\documentclass{article}
\usepackage[T1]{fontenc}
\usepackage[utf8]{inputenc}
\usepackage{ismir}
\usepackage{amsmath,cite,url}
\usepackage{graphicx}
\usepackage{color}
\usepackage{booktabs}
\usepackage[font=small,skip=2pt]{caption}

\title{Do Joint Language-Audio Embeddings Encode Perceptual Timbre Semantics?}

 \oneauthor
  {Qixin Deng \qquad Bryan Pardo \qquad Thrasyvoulos N. Pappas}
  {Northwestern University \\
   \texttt{qixindeng2026@u.northwestern.edu}}

\def\authorname{Q. Deng, B. Pardo, and T. N. Pappas}

\begin{document}

\maketitle

\begin{abstract}
Understanding and modeling the relationship between language and sound are essential for applications such as music information retrieval, text-guided music generation, and audio captioning. Central to these tasks are joint language-audio embedding spaces, which map textual descriptions and auditory content into a shared representation. Although multimodal embedding models such as MS-CLAP, LAION-CLAP, MuQ-MuLan, and OpenFLAM have shown strong performance in language-audio alignment, their correspondence to human perception of timbre, a multifaceted attribute encompassing qualities such as brightness, roughness, and warmth, remains under-explored. In this paper, we evaluate these joint language-audio embedding models in terms of their ability to capture perceptual timbre semantics. Across two complementary experiments, we find that LAION-CLAP shows relatively strong and consistent alignment with human-perceived timbre semantics across both instrumental sounds and descriptor-conditioned audio effects. At the same time, the overall strength of this alignment remains limited, suggesting that current joint language-audio embeddings capture perceptual timbre semantics only partially. We also observe that, overall, reverb-induced timbre semantics are more consistently encoded than equalization-induced timbre semantics.
\end{abstract}

\section{Introduction}\label{sec:introduction}

Joint language–audio embeddings map text and sound into a shared space, bringing semantically related pairs closer together and enabling tasks such as cross-modal retrieval\cite{deshmukh23_interspeech}, audio captioning\cite{10446672}, text-guided audio effects \cite{10890334}, and music generation \cite{pmlr-v202-liu23f}. Recent embedding models, such as MS-CLAP \cite{CLAP2022, 10448504}, LAION-CLAP \cite{laionclap2023}, MuQ-MuLan \cite{zhu2025muq} and OpenFLAM \cite{flam2025} perform well at identifying content (e.g., saxophone solos, footsteps). Less clear is whether they capture more subtle perceptual qualities of timbre \cite{lemaitre2019timbre, mcadams2019perceptual}. A saxophone may be warm, bright, or raspy, while footsteps might sound light, crunchy, or heavy, attributes often subtle and underrepresented in training metadata.

Research on timbral semantics has followed two paths. One examines instruments: Jiang et al. \cite{8940168} distilled 329 descriptors into 16 core terms (e.g., bright–dark, raspy–mellow), while Roche et al. \cite{roche2021metallic} clustered 784 expressions from French listeners into eight perceptual dimensions. The other links descriptors to audio effects: SocialFX \cite{zheng2016socialfx} crowdsourced EQ, reverb, and compression terms from 480+ participants, yielding hundreds of descriptors. Across studies, adjectives like warm, bright, sharp, and clear consistently recur, suggesting perceptual patterns generalize across sources and effects.

Recent work by Tian et al. \cite{tian2025assessing} showed that CLAP-derived representations achieve strong alignment with human timbre similarity judgments, outperforming many conventional audio representations. However, their evaluation focuses on audio-only similarity rather than semantic relationships. Only a few studies have examined more directly how language-audio embeddings represent timbre-related semantics. While Text2FX \cite{10890334} explores whether MS-CLAP encodes timbral semantics at all, it does not evaluate the extent or consistency of this encoding. Deng et al. \cite{deng2025investigating} investigate the sensitivity of LAION-CLAP to common audio effects such as gain, low-pass filtering, reverberation, and bitcrushing, showing that its embeddings respond systematically to effect strength. Hebbar et al. \cite{hebbar2025investigating} study timbre representations in LAION-CLAP across audio and text modalities by identifying perturbation directions for adjectives such as \emph{bright}, \emph{dark}, and \emph{distorted}. Their results provide evidence for a shared timbre-related subspace across modalities, but their analysis is limited to a single model and a small set of descriptors.

In this work, we assess the perceptual validity of joint language–audio embeddings with respect to timbral semantics. We use human-annotated datasets from Jiang et al. \cite{8940168} and connect audio effects to language through descriptor-conditioned manipulations generated. with SocialFX \cite{zheng2016socialfx}. On this data we evaluate how popular embedding spaces preserve timbral characteristics and how this preservation varies across models. 

The contributions of our work are: 1). A methodology for assessing the alignment of language-audio embedding models with human perception of timbre. 2). An evaluation and comparison between popular language-audio embeddings using this methodology.

\section{The Models} 
\label{subsec:the-models}
We focus on four language–audio embedding models: MS-CLAP \cite{CLAP2022, 10448504}, LAION-CLAP \cite{laionclap2023}, MuQ-MuLan \cite{zhu2025muq} and OpenFLAM \cite{flam2025}. These differ in training data and domain coverage and are representative of the range of language-audio models likely to be used in current systems.

MS-CLAP and LAION-CLAP are widely-used models that target general audio understanding, meaning that they are trained to represent a broad spectrum of sounds, including music, speech, environmental sounds(e.g., dogs barking, doors closing, waves crashing) and abstract auditory events (e.g., alarms, sirens). MS-CLAP is trained on a combination of datasets, including FSD50k, Clotho V2, AudioCaps, and MACS. These span music, speech, natural sounds, and abstract auditory events paired with human-written captions. LAION-CLAP uses their own curated large-scale LAION-Audio-630k dataset, which contains environmental and human-related audio clips labeled via keyword-to-caption augmentation. Both models are trained on clip-level labels and are not trained to provide fine temporal granularity. MuQ-MuLan, focuses specifically on music and is trained on video soundtracks paired with metadata such as tags, titles, descriptions, etc. 

OpenFLAM is a language-audio embedding model with both clip-level and frame-level representations, designed for finer-grained audio-text alignment. In our experiments, we use its clip-level global representation because the audio samples used in this study are short clips with relatively consistent content throughout the duration, we consider the global representation more appropriate for capturing the overall timbral characteristics of each sample.

\section{Experiments}
\label{sec:exp}
We performed two experiments to evaluate the alignment between the four audio-text embedding models (see Section \ref{subsec:the-models}) and human perception of timbre. In the first experiment, we assessed whether language-audio embedding models capture human-perceived timbre relationships on musical instruments. In the second experiment, we investigated how these embedding models capture perceptual timbre descriptors in relation to changes in audio effects, specifically equalization (EQ) and reverberation.

\subsection{Experiment 1: Instrumental Timbre Semantics}
In the first experiment, we assessed whether language-audio embedding models capture human-perceived timbral semantics at both the descriptor and instrument level, using Jiang’s CCMusic-Database-Instrument-Timbre dataset. 

\subsubsection{Instrument Timbre Dataset}
\begin{figure}
    \centering
    \includegraphics[width=1.0\linewidth]{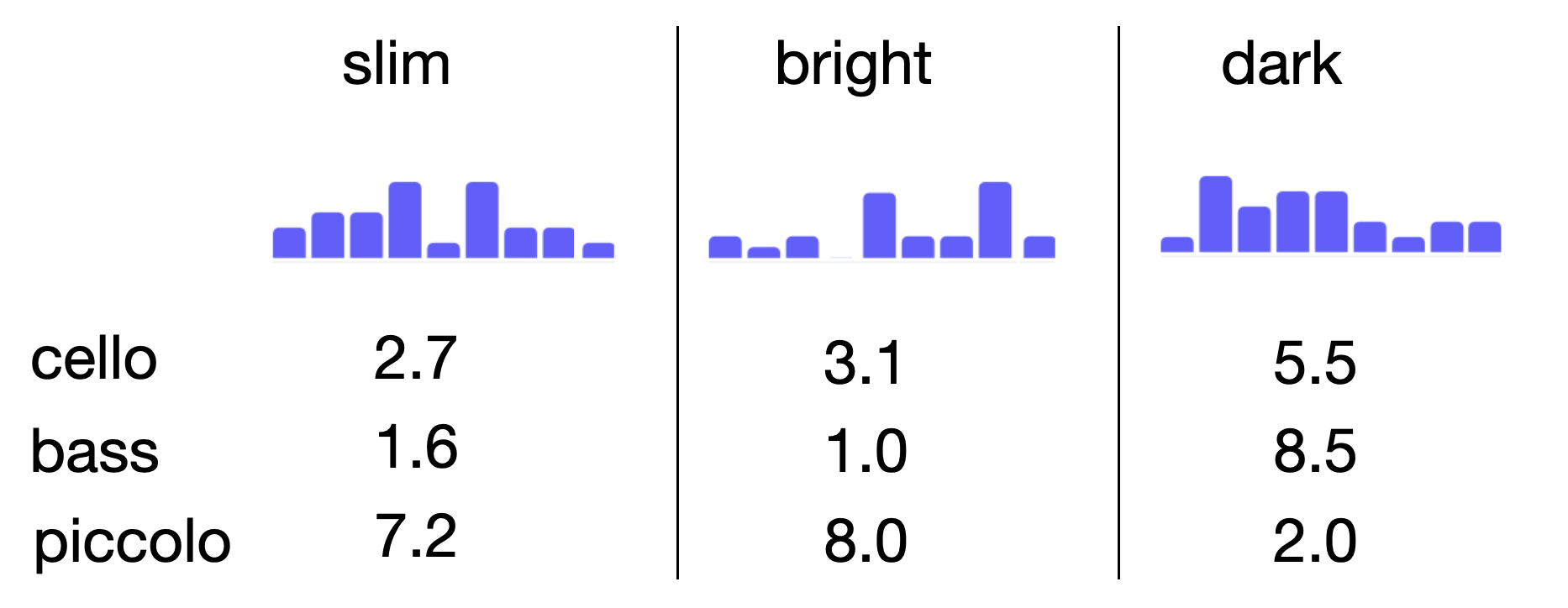}
    \caption{An illustrative exercpt of Jiang's CCMusic-Database-Instrument-Timbre dataset, showing three descriptors and three Western instruments. The bargraph under each descriptor shows the distribution of ratings for that descriptor, over all rated Western instrument. The numbers in the table show  mean values returned by listeners for the given descriptor-instrument pair.  The full table has 61 instrument rows and 16 descriptor columns.}
    \label{fig:human-data}
\end{figure}

Our human judgements of musical instrument timbre come from Jiang’s CCMusic-Database-Instrument-Timbre dataset\cite{8940168}. This dataset is openly available and was obtained through a controlled listening test, providing a reliable ground truth for perceptual timbre semantics. The dataset contains short clips of 37 Chinese and 24 Western instruments, sampled at a rate of 44,100 Hz.  Each clip was rated by 34 musically trained listeners (16 male, 18 female) on a 9-point scale for its correspondence to each of 16 descripive terms (e.g., bright, dark, raspy). See Figure \ref{fig:descriptor} for the full list of descriptors. An illustrative excerpt of this data is shown in Figure \ref{fig:human-data}.

\subsubsection{Encoding the data}
 
For each embedding model, we encoded each instrument audio clip to obtain an audio embedding $\mathbf{a}_i$. We similarly encoded each text descriptor to obtain a text embedding $\mathbf{t}_d$. Cosine similarity was computed between each audio embedding and each text embedding, yielding a 61 (instrument) by 16 (descriptor) matrix of embedding ratings $E$. Each entry $E_{i,d}$ reflects the strength of association, in the joint embedding space, between instrument $i$ and descriptor $d$, as measured by cosine similarity. 

Let $H$ be the matrix of mean human ratings, as illustrated in Figure \ref{fig:human-data}. Here, $H_{i,d}$ is the mean human rating for instrument $i$ on descriptor $d$ As with the embedding data, this is a 61 (instrument) by 16 (descriptor) matrix.

 \subsubsection{Descriptor-level correlation}
 
 We compare the human ratings $H$ to the embedding ratings $E$ using Pearson correlation. This allows us to avoid scaling issues and focus on trends. 
 
 To evaluate a given descriptor $d$, we correlate all elements in column $d$ of the matrices: $H_{*,d}$ and $E_{*,d}$.  High positive correlation 
 reflects semantic alignment for that perceptual quality. A low correlation suggests weak alignment between the embedding space and human perception for that descriptor. A negative correlation indicates a mismatch, where instruments rated highly on descriptor $d$ by humans are placed \emph{farther away} from the descriptor in the embedding space, suggesting the model encodes an opposite or contradictory association.

\subsubsection{2. Instrument-level semantic profile correlation} For a given embedding model, we can also aggregate data by instrument, instead of descriptor. To evaluate a given instrument $i$, we correlate all elements in row $i$ of the matrices: $H_{i,*}$ and $E_{i,*}$. A high correlation indicates that the embedding captures the overall timbre profile of the instrument $i$ across descriptors. A low correlation implies that the embedding fails to reproduce the joint configuration of timbral attributes as perceived by listeners. A negative correlation indicates a systematic inversion, where descriptors that listeners strongly associate with an instrument are those that the embedding places far away, suggesting the model misrepresents the instrument’s perceptual timbre profile.

\subsubsection{Results for Experiment 1:}

\begin{figure}[htb]
    \begin{minipage}[b]{1.0\linewidth}
     \centering
     \includegraphics[width=1.0\textwidth]{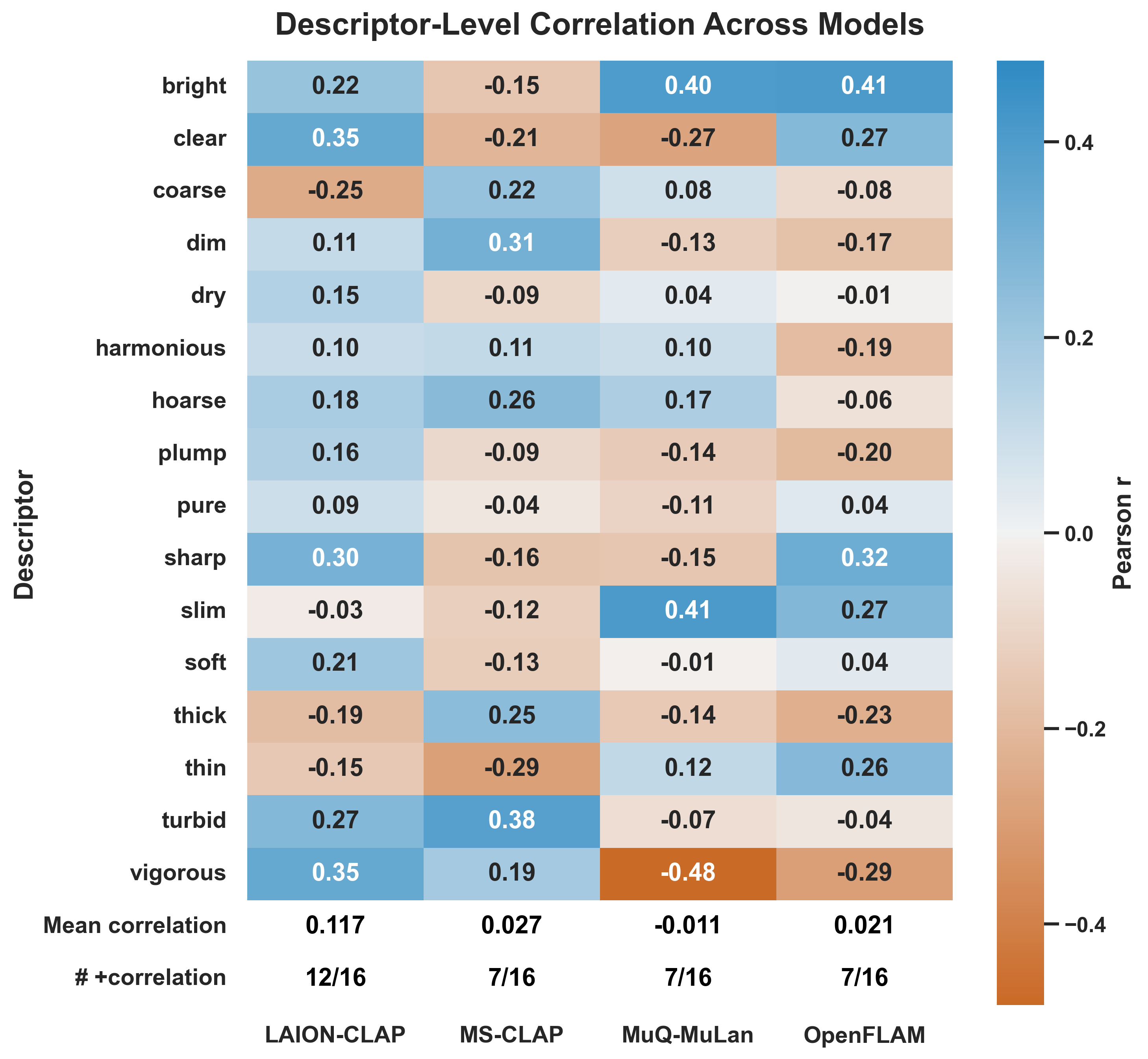}
     \caption{Descriptor-level Pearson correlations between human timbre ratings and model-based language-audio similarities across the four evaluated models.}
     \label{fig:descriptor}
     \end{minipage}
  
\end{figure}

At the descriptor level, LAION-CLAP demonstrated the strongest alignment with human ratings, with 12 out of 16 descriptors showing positive correlations and the highest overall mean correlation ($r = 0.117$). In contrast, MS-CLAP achieved positive correlations for only 7 of the 16 descriptors, with a much lower mean correlation ($r = 0.027$). MuQ-MuLan likewise produced positive correlations for 7 descriptors, but its overall mean correlation was slightly negative ($r = -0.011$), indicating weak descriptor-level alignment overall. OpenFLAM also showed 7 positive correlations out of 16 descriptors, with a small positive mean correlation ($r = 0.021$), placing it slightly above MuQ-MuLan but still substantially below LAION-CLAP. Overall, these results indicate that LAION-CLAP provides the most consistent descriptor-level alignment with human timbre perception among the four models, as visually demonstrated in the descriptor-level correlation heatmap in Fig.~\ref{fig:descriptor} .

At the instrument level, LAION-CLAP again demonstrated the strongest alignment for Chinese instruments, yielding 24 positive correlations out of 37 instruments with the highest mean profile correlation ($r = 0.162$). MS-CLAP produced the same number of positive correlations (24/37), but with a notably weaker average correlation ($r = 0.058$). OpenFLAM followed closely, with 23 positive correlations out of 37 and a mean correlation of $r = 0.069$, indicating moderate but still weaker alignment than LAION-CLAP. MuQ-MuLan was the least consistent for Chinese instruments, with only 16 positive correlations and a mean correlation of approximately zero, suggesting little overall alignment. 

\begin{figure}[htb]
    \begin{minipage}[b]{1.0\linewidth}
     \centering
     \includegraphics[width=1.0\textwidth]{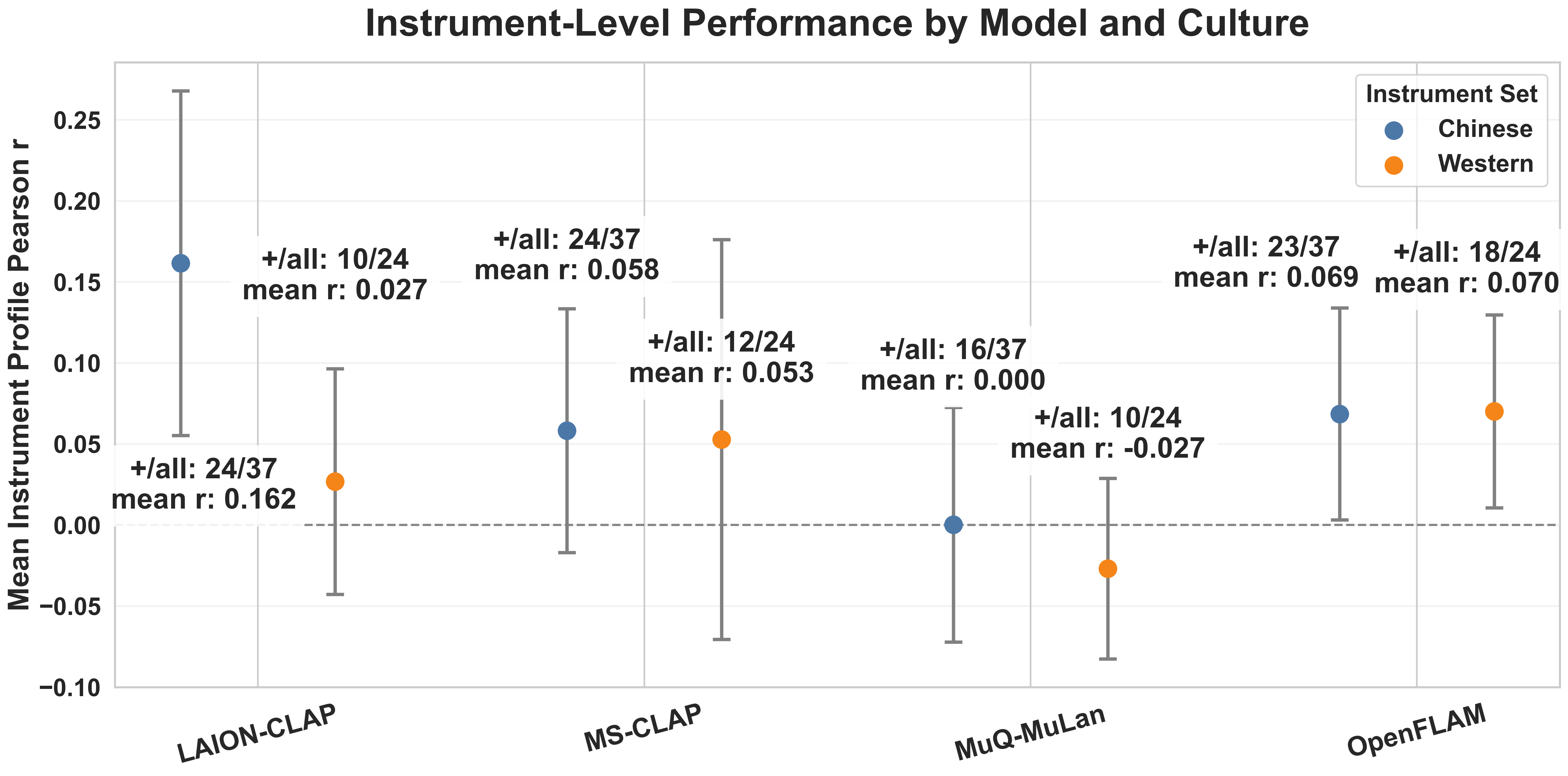}
     \caption{Instrument-level semantic profile correlation results for the four evaluated models for both Chinese and Western instruments. Each point denotes the mean Pearson correlation between the human descriptor rating profile and the embedding-based similarity profile across instruments within that group. Error bars indicate variability across instruments.}
     \label{fig:instrument}
     \end{minipage}
  
\end{figure}

For Western instruments, OpenFLAM achieved the strongest performance, producing 18 positive correlations out of 24 instruments, substantially exceeding the other three models. MS-CLAP showed the next best alignment, with 12 positive and 12 negative correlations and a mean correlation of $r = 0.053$. LAION-CLAP was weaker on the Western subset, with 10 positive and 14 negative correlations and a mean correlation of $r = 0.027$. MuQ-MuLan produced the same number of positive correlations as LAION-CLAP (10/24), but its mean correlation was slightly negative ($r = -0.027$), indicating the weakest performance overall on Western instruments. As shown in Fig.~\ref{fig:instrument}, these results suggest that LAION-CLAP captures instrumental timbre semantics most reliably overall, particularly for descriptor-level analysis and Chinese instruments, while OpenFLAM shows a relative advantage on Western instrument-level semantic profiles. That said, no model showed strong across-the-board correlation with humans.

\subsection{Experiment 2: Audio Effect Timbre Semantics}

While Experiment 1 evaluates embeddings using naturally occurring timbral variation across instruments, such recordings also vary in pitch, dynamics, and recording conditions, making it difficult to isolate timbre alone. Experiment 2 therefore uses controlled digital signal processing (DSP) manipulations to vary timbre in a more systematic manner. This design builds on SocialFX~\cite{zheng2016socialfx}, a large crowdsourced dataset linking 4,297 unique vocabulary terms to quantified audio effect parameter settings. These mappings provide a perceptually grounded reference for how layperson descriptors (e.g., \emph{warm}, \emph{harsh}) correspond to measurable timbral changes. Two effect types were considered:

\textbf{1. Equalization (EQ):} implemented using a 40-band parametric equalizer, where each band is defined by a center frequency, bandwidth, and gain. For each descriptor, SocialFX specifies a gain profile across the 40 bands. An amount scaling factor was applied to linearly scale all band gains and control the strength of the EQ manipulation.

\textbf{2. Reverberation:} implemented using a digital reverberator consisting of parallel comb filters, all-pass filters, and low-pass filters. Parameters included decay time, feedback gain, modulation, low-pass cutoff frequency, and overall effect gain. Effect strength was controlled through the wet/dry mix.

We selected the 20 most frequently used EQ descriptors and the 20 most frequently used reverb descriptors from the SocialFX vocabulary. For each descriptor, manipulated audio was generated using descriptor-specific effect settings derived from SocialFX. The manipulations were applied to three source audio signals: guitar, piano, and a classical ensemble excerpt. We used the DSP implementations provided by the Audealize website~\cite{seetharaman2016audealize}.

For each descriptor-source pair, manipulations were generated at five effect strengths: $0.2$, $0.4$, $0.6$, $0.8$, and $1.0$. For EQ, these values scaled the descriptor-specific gain profile. For reverb, they controlled the wet/dry mix.

For each embedding model, the following steps were performed:

\textbf{1. Text embeddings.} A text embedding was computed for each SocialFX descriptor $d$.

\textbf{2. Audio embeddings.} For each source audio $a$ and descriptor $d$, audio embeddings were computed for the original source signal and for the manipulated versions obtained by applying a single effect type (EQ or reverb) at each of the five effect strengths.

\textbf{3. Similarity computation.} Cosine similarity in the joint embedding space, $sim(d, \cdot)$, was computed between the descriptor embedding and each corresponding audio embedding.

We define the change in similarity due to manipulation as
$$
\Delta_{d,e,l,a} = sim(d, fx(a,e,l)) - sim(d,a)
$$
where $fx(\cdot)$ applies effect $e$ (EQ or reverb) at level $l$ to source audio $a$. A positive $\Delta_{d,e,l,a}$ indicates that the manipulation moves the audio embedding closer to the descriptor $d$ in the joint embedding space.

\subsubsection{Analysis and Results for Experiment 2:}
\begin{table*}[t]
\centering
\caption{Model-level summary statistics for encoding audio effect induced timbre transformations.}
\label{tab:model_summary}
\resizebox{\textwidth}{!}{%
\begin{tabular}{lccccc}
\toprule
Model & Pos. slope rate $\uparrow$ & Pos. final-$\Delta$ rate $\uparrow$ & All-source consistency $\uparrow$ & Mean Pearson $r$ $\uparrow$ & Mean Spearman $\rho$ $\uparrow$ \\
\midrule
LAION-CLAP & 72.5\% & 77.5\% & 42.5\% & 0.354 & 0.334 \\
MS-CLAP     & 71.7\% & 75.8\% & 42.5\% & 0.361 & 0.347 \\
MuQ-MuLan  & 78.3\% & 64.2\% & 47.5\% & 0.535 & 0.530 \\
OpenFLAM   & 52.5\% & 40.0\% & 22.5\% & 0.069 & 0.073 \\
\bottomrule
\end{tabular}%
}
\end{table*}
\begin{table*}[t]
\centering
\caption{Summary of model performance by effect type.}
\label{tab:effect_summary}
\resizebox{\textwidth}{!}{%
\begin{tabular}{llccccc}
\toprule
Effect & Model & Pos. slope rate (\%) $\uparrow$ & Pos. final-$\Delta$ rate (\%) $\uparrow$ & All-source consistency (\%) $\uparrow$ & Mean Pearson $r$ $\uparrow$ & Mean Spearman $\rho$ $\uparrow$ \\
\midrule
eq     & LAION-CLAP & 61.7\% & 63.3\% & 30.0\% & 0.186  & 0.148 \\
eq     & MS-CLAP     & 73.3\% & 71.7\% & 40.0\% & 0.336  & 0.300 \\
eq     & MuQ-MuLan  & 68.3\% & 55.0\% & 30.0\% & 0.396  & 0.408 \\
eq     & OpenFLAM   & 56.7\% & 40.0\% & 45.0\% & 0.139  & 0.120 \\
reverb & LAION-CLAP & 83.3\% & 91.7\% & 55.0\% & 0.522  & 0.520 \\
reverb & MS-CLAP     & 70.0\% & 80.0\% & 45.0\% & 0.386  & 0.393 \\
reverb & MuQ-MuLan  & 88.3\% & 73.3\% & 65.0\% & 0.674  & 0.652 \\
reverb & OpenFLAM   & 48.3\% & 40.0\% & 0.0\%  & -0.001 & 0.027 \\
\bottomrule
\end{tabular}%
}
\end{table*}

To evaluate each model’s ability to encode audio-effect-induced perceptual timbre semantics, we consider four aspects: \textbf{overall trend}, \textbf{cross-source robustness}, \textbf{linearity}, and \textbf{monotonicity}.

\textbf{1. Overall trend.} Overall trend is characterized using the slope of each descriptor-effect-source trajectory and the final similarity change at the strongest manipulation level. A positive slope indicates that similarity to the target descriptor tends to increase as effect strength increases. A positive final $\Delta$ indicates that the most strongly manipulated audio is closer to the target descriptor than the original audio. At the model level, these are summarized by the \emph{positive slope rate} and the \emph{positive final-$\Delta$ rate}, which measure the proportion of trajectories with positive slope and positive final $\Delta$, respectively.

\textbf{2. Cross-source robustness.} Cross-source robustness evaluates whether a descriptor-aligned trend generalizes across different source signals rather than depending on a particular audio input. We quantify this using \emph{all-source consistency}, defined as the proportion of descriptor-effect pairs for which the slope is positive for all source types.

\textbf{3. Linearity.} Linearity characterizes how smoothly target-descriptor similarity changes with effect strength. We quantify this using the Pearson correlation coefficient $r$ between manipulation level and $\Delta_{d,e,l,a}$, and report the average Pearson $r$ across all descriptor-effect-source trajectories for each model.

\textbf{4. Monotonicity.} Monotonicity characterizes whether target-descriptor similarity tends to increase consistently as effect strength increases, even when the relationship is not perfectly linear. We quantify this using the Spearman rank correlation coefficient $\rho$ between manipulation level and $\Delta_{d,e,l,a}$, and report the average Spearman $\rho$ across all trajectories for each model.

As shown in Table~\ref{tab:model_summary}, MuQ-MuLan achieves the strongest performance on most metrics, with the highest positive slope rate (78.3\%), the highest all-source consistency (47.5\%), and the largest mean Pearson $r$ (0.535) and mean Spearman $\rho$ (0.530). This indicates that its embedding trajectories most consistently move in the intended descriptor direction and do so with the strongest linear and monotonic structure. LAION-CLAP and MSCLAP show similar behavior in terms of overall trend and cross-source robustness. LAION-CLAP attains a positive slope rate of 72.5\% and an all-source consistency of 42.5\%, while MSCLAP achieves 71.7\% and 42.5\%, respectively. However, LAION-CLAP yields the highest positive final-$\Delta$ rate (77.5\%), slightly above MSCLAP (75.8\%), indicating that it most often places the most strongly manipulated audio closer to the target descriptor at the endpoint. OpenFLAM performs substantially worse on all metrics, with a positive slope rate of 52.5\%, a positive final-$\Delta$ rate of 40.0\%, and very low mean Pearson and Spearman correlations, suggesting weaker sensitivity.

We further analyze performance separately for the two effect types, EQ and reverb. As shown in Table~\ref{tab:effect_summary}, reverb generally yields stronger descriptor-aligned embedding trends than EQ for most models. In particular, LAION-CLAP and MuQ-MuLan both show markedly higher positive slope rates and substantially stronger Pearson and Spearman correlations for reverb than for EQ, indicating that reverb-induced timbre transformations are tracked more reliably and with a clearer trajectory structure in the embedding space. MSCLAP shows a weaker effect-type contrast, although its correlation and endpoint metrics are also stronger for reverb. OpenFLAM performs poorly for both effect types. Taken together, these results indicate that reverb manipulations tend to produce more readily encoded timbre-semantic changes than EQ manipulations.

\begin{figure}[t]
    \centering
    \includegraphics[width=0.9\columnwidth]{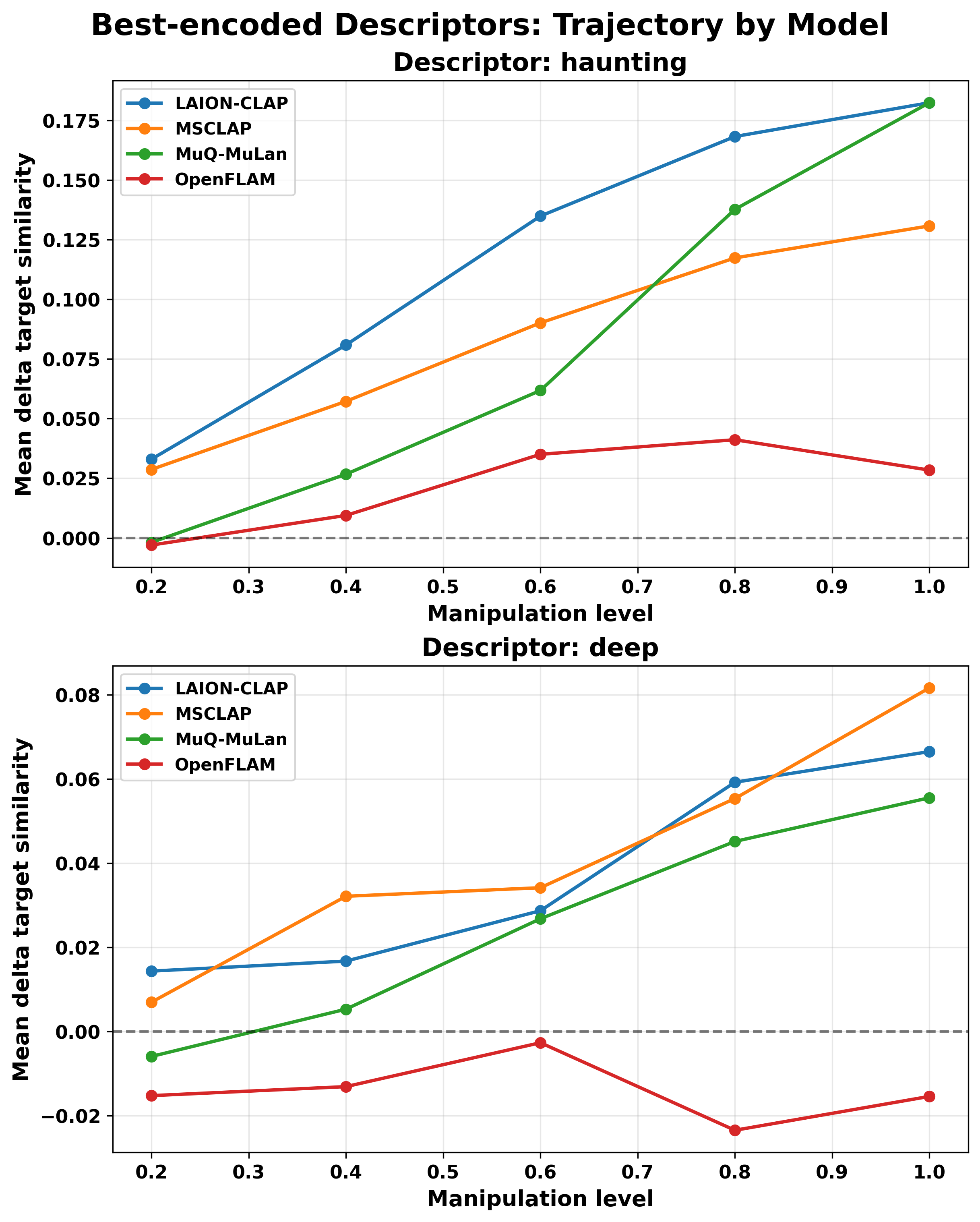}
    \caption{Movement trajectories for the two most robustly encoded descriptors, \emph{haunting} and \emph{deep}, across embedding models. Each curve shows the change in target-descriptor similarity, $\Delta$, as manipulation strength increases. The consistently positive trajectories for LAION-CLAP and MuQ-MuLan, together with the weaker response of OpenFLAM, support the model-level findings in Table~\ref{tab:model_summary}.}
    \label{fig:best_encode}
\end{figure}

In addition, we identified descriptors whose encoding is relatively robust by aggregating results across all models, effect types, and source signals. Here, a descriptor is regarded as well encoded if stronger descriptor-driven manipulation generally moves the audio embedding in the intended semantic direction, if this behavior appears in a large proportion of conditions, and if it is not limited to only one source signal. We therefore first computed condition-level trajectory statistics for each $(\text{model}, \text{effect type}, \text{source type}, \text{descriptor})$ combinations, and then summarized these statistics at the descriptor level. Descriptors were ranked according to their average directional behavior, the frequency with which they showed positive trends, and their consistency across source types. Using this procedure, the descriptors identified as the most robustly encoded were \emph{haunting}, \emph{deep}, \emph{church}, \emph{spacious}, \emph{distorted}, and \emph{echo}. Notably, these descriptors are primarily associated with reverberation-related transformations, suggesting that such timbre semantics are encoded more reliably than many EQ-related descriptors. The movement trajectories of the two strongest descriptors, \emph{haunting} and \emph{deep}, across models are shown in Fig.~\ref{fig:best_encode}. These trajectories further support two main findings: LAION-CLAP and MuQ-MuLan provide the strongest encoding while OpenFLAM provides the weakest, and reverb-related descriptors produce the most consistent descriptor-aligned movement in the evaluated embedding spaces.

\section{Conclusion and Future Work}

We conducted a systematic evaluation of four prominent joint language-audio embedding models: MS-CLAP, LAION-CLAP, MuQ-MuLan, and OpenFLAM. While all evaluated models showed only limited alignment with human-perceived timbre semantics, LAION-CLAP demonstrated relatively stronger and more consistent alignment across experiments. This may be related to its large-scale and diverse language-audio pretraining, which may expose the model to broader associations between auditory characteristics, including timbre-related attributes, and descriptive language. We also found that reverb-induced timbre semantics were generally encoded more robustly than EQ-induced semantics. Overall, these results suggest that current joint language-audio embeddings struggle to fully capture perceptual timbre semantics.

Future work may proceed in two directions. First, examining whether LAION-CLAP encodes interpretable timbral axes, such as a perceptual continuum from \emph{bright} to \emph{dark}, could provide deeper insight into the structure of its embedding space and its relationship to perceptual dimensions of timbre. Second, fine-tuning LAION-CLAP with timbre-specific objectives may improve its ability to capture subtle timbral qualities, thereby supporting more effective timbre-based retrieval, manipulation, and generative applications.

\section{AI Usage Statement}

Large language models were used only to assist with language editing, including correcting grammar, improving sentence clarity, and refining wording.

\bibliography{ISMIRtemplate}

@incollection{mcadams2019perceptual,
  title={The perceptual representation of timbre},
  author={McAdams, Stephen},
  booktitle={Timbre: Acoustics, perception, and cognition},
  pages={23--57},
  year={2019},
  publisher={Springer}
}

@incollection{lemaitre2019timbre,
  title={Timbre, sound quality, and sound design},
  author={Lemaitre, Guillaume and Susini, Patrick},
  booktitle={Timbre: Acoustics, perception, and cognition},
  pages={245--272},
  year={2019},
  publisher={Springer}
}

@inproceedings{deshmukh23_interspeech,
  title     = {{Audio Retrieval with WavText5K and CLAP Training}},
  author    = {Soham Deshmukh and Benjamin Elizalde and Huaming Wang},
  year      = {2023},
  booktitle = {{Interspeech 2023}},
  pages     = {2948--2952},
  doi       = {10.21437/Interspeech.2023-1136},
  issn      = {2958-1796},
}

@INPROCEEDINGS{10446672,
  author={Kim, Jaeyeon and Jung, Jaeyoon and Lee, Jinjoo and Woo, Sang Hoon},
  booktitle={ICASSP 2024 - 2024 IEEE International Conference on Acoustics, Speech and Signal Processing (ICASSP)}, 
  title={EnCLAP: Combining Neural Audio Codec and Audio-Text Joint Embedding for Automated Audio Captioning}, 
  year={2024},
  volume={},
  number={},
  pages={6735-6739},
  doi={10.1109/ICASSP48485.2024.10446672}}

@inproceedings{CLAP2022,
  title={Clap learning audio concepts from natural language supervision},
  author={Elizalde, Benjamin and Deshmukh, Soham and Al Ismail, Mahmoud and Wang, Huaming},
  booktitle={ICASSP 2023-2023 IEEE International Conference on Acoustics, Speech and Signal Processing (ICASSP)},
  pages={1--5},
  year={2023},
  organization={IEEE}
}

@INPROCEEDINGS{10448504,
  author={Elizalde, Benjamin and Deshmukh, Soham and Wang, Huaming},
  booktitle={ICASSP 2024 - 2024 IEEE International Conference on Acoustics, Speech and Signal Processing (ICASSP)}, 
  title={Natural Language Supervision For General-Purpose Audio Representations}, 
  year={2024},
  volume={},
  number={},
  pages={336-340},
  doi={10.1109/ICASSP48485.2024.10448504}}

@inproceedings{laionclap2023,
  title = {Large-scale Contrastive Language-Audio Pretraining with Feature Fusion and Keyword-to-Caption Augmentation},
  author = {Wu, Yusong and Chen, Ke and Zhang, Tianyu and Hui, Yuchen and Berg-Kirkpatrick, Taylor and Dubnov, Shlomo},
  booktitle={IEEE International Conference on Acoustics, Speech and Signal Processing, ICASSP},
  year = {2023}
}

@article{zhu2025muq,
      title={MuQ: Self-Supervised Music Representation Learning with Mel Residual Vector Quantization}, 
      author={Haina Zhu and Yizhi Zhou and Hangting Chen and Jianwei Yu and Ziyang Ma and Rongzhi Gu and Yi Luo and Wei Tan and Xie Chen},
      journal={arXiv preprint arXiv:2501.01108},
      year={2025}
}

@inproceedings{flam2025,
    title={{FLAM}: Frame-Wise Language-Audio Modeling},
    author={Yusong Wu and Christos Tsirigotis and Ke Chen and Cheng-Zhi Anna Huang and Aaron Courville and Oriol Nieto and Prem Seetharaman and Justin Salamon},
    booktitle={Forty-second International Conference on Machine Learning (ICML)},
    year={2025},
    url={https://openreview.net/forum?id=7fQohcFrxG}
}

@INPROCEEDINGS{8940168,
  author={Jiang, Wei and Liu, Jingyu and Li, Zijin and Zhu, Jiaxing and Zhang, Xiaoyi and Wang, Shuang},
  booktitle={2019 IEEE/ACIS 18th International Conference on Computer and Information Science (ICIS)}, 
  title={Analysis and Modeling of Timbre Perception Features of Chinese Musical Instruments}, 
  year={2019},
  volume={},
  number={},
  pages={191-195},
  doi={10.1109/ICIS46139.2019.8940168}}

@inproceedings{zheng2016socialfx,
  author    = {Zheng, Tianran  and Seetharaman,Prem and Pardo, Bryan},
  title     = {Socialfx: Studying a Crowdsourced Folksonomy of Audio Effects Terms},
  booktitle = {Proceedings of the 24th ACM International Conference on Multimedia (ACM MM)},
  year      = {2016},
  pages     = {182--186},
  publisher = {ACM}
}

@article{seetharaman2016audealize,
  author  = {Seetharaman,Prem and Pardo,Bryan},
  title   = {Audealize: Crowdsourced Audio Production Tools},
  journal = {Journal of the Audio Engineering Society},
  volume  = {64},
  number  = {9},
  pages   = {683--695},
  year    = {2016}
}

@article{roche2021metallic,
  author  = {Roche, Fanny and Hueber, Thomas and Garnier, Maëva and  Limier, Samuel and Girin, Laurent},

  title   = {Make That Sound More Metallic: Towards a Perceptually Relevant Control of the Timbre of Synthesizer Sounds Using a Variational Autoencoder},
  journal = {Transactions of the International Society for Music Information Retrieval},
  volume  = {4},
  number  = {1},
  pages   = {115--131},
  year    = {2021},
  doi     = {10.5334/tismir.76},
  url     = {https://transactions.ismir.net/articles/10.5334/tismir.76}
}

@InProceedings{pmlr-v202-liu23f,
  title = 	 {{A}udio{LDM}: Text-to-Audio Generation with Latent Diffusion Models},
  author =       {Liu, Haohe and Chen, Zehua and Yuan, Yi and Mei, Xinhao and Liu, Xubo and Mandic, Danilo and Wang, Wenwu and Plumbley, Mark D},
  booktitle = 	 {Proceedings of the 40th International Conference on Machine Learning},
  pages = 	 {21450--21474},
  year = 	 {2023},
  editor = 	 {Krause, Andreas and Brunskill, Emma and Cho, Kyunghyun and Engelhardt, Barbara and Sabato, Sivan and Scarlett, Jonathan},
  volume = 	 {202},
  series = 	 {Proceedings of Machine Learning Research},
  month = 	 {23--29 Jul},
  publisher =    {PMLR},
  url = 	 {https://proceedings.mlr.press/v202/liu23f.html}
}

@INPROCEEDINGS{10890334,
  author={Chu, Annie and O’Reilly, Patrick and Barnett, Julia and Pardo, Bryan},
  booktitle={ICASSP 2025 - 2025 IEEE International Conference on Acoustics, Speech and Signal Processing (ICASSP)}, 
  title={Text2FX: Harnessing CLAP Embeddings for Text-Guided Audio Effects}, 
  year={2025},
  volume={},
  number={},
  pages={1-5},
  doi={10.1109/ICASSP49660.2025.10890334}}

@inproceedings{tian2025assessing,
  title     = {Assessing the Alignment of Audio Representations with Timbre Similarity Ratings},
  author    = {Tian, Haokun and Lattner, Stefan and Saitis, Charalampos},
  booktitle = {Proceedings of the 26th International Society for Music Information Retrieval Conference (ISMIR)},
  year      = {2025},
  address   = {Daejeon, South Korea},
  publisher = {International Society for Music Information Retrieval},
}

@inproceedings{
hebbar2025investigating,
title={Investigating Timbre Representations in {CLAP} Across Modalities via Perturbations},
author={Devyani Hebbar and Brian McFee},
booktitle={1st International Workshop on Emerging AI Technologies for Music},
year={2025},
url={https://openreview.net/forum?id=KKzcDulLRm}
}

@article{deng2025investigating,
  title = {Investigating the Sensitivity of Pre-trained Audio Embeddings to Common Effects},
  author = {Deng, Victor and Wang, Changhong and Richard, Ga{\"e}l and McFee, Brian},
  journal = {IEEE International Conference on Acoustics and Speech and and Signal Processing (ICASSP)},
  year = {2025},
}

%
%
%
%

\end{document}